\documentclass{PoS}

\newcommand{\fig}{Fig.\,}

\title{The i-process and CEMP-r/s stars}

\ShortTitle{The i-process and CEMP-r/s stars}

\author{L. Dardelet$^a$, C. Ritter$^{b,c,d}$, P. Prado$^b$, E. Heringer$^b$, C. Higgs$^b$, S. Sandalski$^{h,c}$, S.~Jones$^{b,d}$, 
P.~Denissenkov$^{b,c,d}$, K.Venn$^b$, M. Bertolli$^{e,f,d}$, M. Pignatari$^{g,d}$, P. Woodward$^{h,c}$, F.~Herwig$^{b,c,d}$ \\
\llap{$^a$}Department of Physics, \'{E}cole Normale Sup\'{e}rieure, 45 rue d'Ulm, 75005 Paris, France
\llap{$^b$}Department of Physics and Astronomy, University of Victoria, 
Victoria, BC, V8P5C2, Canada \\
\llap{$^c$}Joint Institute for Nuclear Astrophysics, Center for the Evolution of the Elements, Michigan State University,
640 South Shaw Lane, East Lansing, MI 48824, USA\\
\llap{$^d$}NuGrid collaboration, http://www.nugridstars.org\\
\llap{$^e$}Oak Ridge National Laboratory, 
P.O. Box 2008,Oak Ridge, TN 37831, USA\\
\llap{$^f$}Department of Physics and Astronomy, University of Tennessee, 
Knoxville, TN 37996, USA\\
\llap{$^g$}Department of Physics, University of Basel, 
Klingelbergstrasse 82, CH-4056 Basel, Switzerland\\
\llap{$^h$}LCSE and Department of Astronomy, University of Minnesota, 
Minneapolis, MN 55455, USA\\

E-mail: \email{critter@uvic.ca}}

\abstract{
We investigate whether the anomalous elemental abundance patterns in some of the C-enhanced metal-poor-r/s
(CEMP-r/s) stars 
are consistent with predictions of nucleosynthesis yields from the i-process, a neutron-capture regime at  
neutron densities intermediate between those typical for the slow (s) and rapid (r) processes.
Conditions necessary for the i-process are expected to be met at multiple stellar sites, 
such as the He-core and He-shell flashes in low-metallicity low-mass stars, super-AGB and post-AGB stars, 
as well as low-metallicity massive stars. 
We have found that single-exposure one-zone simulations of the i-process
reproduce the abundance patterns in some of the CEMP-r/s stars much better than
the model that assumes a superposition of yields from s- and r-process sources.
Our previous study of nuclear data uncertainties relevant to the i-process 
revealed that they could have a significant impact on the i-process yields obtained in our idealized one-zone calculations, 
leading, for example, to $\sim 0.7 \mathrm{dex}$ uncertainty in our predicted [Ba/La] ratio. Recent 3D hydrodynamic 
simulations of convection driven by a He-shell flash in post-AGB Sakurai's object have discovered
a new mode of non-radial instabilities: the Global Oscillation of Shell H-ingestion. 
This has demonstrated that spherically symmetric stellar evolution simulations 
cannot be used to accurately model physical conditions for the i-process.
}

\FullConference{XIII Nuclei in the Cosmos,\\
		7-11 July, 2014\\
		Debrecen, Hungary }

\begin{document}

\section{Introduction}


\noindent The fraction of C-Enhanced Metal Poor (CEMP) stars (objects with [C/Fe]>1)
increases with a decrease in metallicity, reaching 20\% at [Fe/H]=-2
\cite{2012A&A...548A..34A,2005ARA&A..43..531B}. 
CEMP-r/s stars form a subclass characterized by simultaneous enhancements of both s- (e.g., Ba) and r-process (e.g., Eu) elements.
\cite{2006A&A...451..651J,2012ApJ...747....2L,2010A&A...509A..93M}.
The prevailing model of a CEMP-r/s star assumes that it is a component of a binary system
initially pre-enriched with r-process material that has additionally accreted s-process material from its heavier 
AGB-star companion, which leads to a superposition of both processes \cite{2006A&A...451..651J}. 
However, this model fails to explain CEMP-r/s stars with large 
[hs/ls]\footnote{hs and ls are the total abundances of selected elements representing the second (high) and 
first (light) s-process peaks in the solar abundance distribution. } and low
[La/Eu] ratios. An example of such a star is CS 31062-050 
shown in 
\fig26 in \cite{2012MNRAS.422..849B}.

\noindent The intermediate neutron-capture process (hereafter, the i-process, see \cite{1977ApJ...212..149C}) operates at 
neutron densities between those characteristic of the s- and r-process,
when H is convectively entrained and advected into a He-burning zone. 
When the advected fluid parcel has reached a depth with $T \approx 1.5\times10^8 \mathrm{K}$,
$^{13}$N is produced in the reaction $^{12}$C(p,$\gamma$)$^{13}$N. 
The beta-decay of $^{13}$N to $^{13}$C followed by the reaction $^{13}$C($\alpha$,n)$^{16}$O
result in a production of neutrons with a density $N_n\approx 10^{15} \mathrm{cm}^{-3}$.
Possible astrophysical sites for the i-process have been identified in
stellar evolution computations. These are the He-shell thermal pulses in the low- and zero-metallicity
AGB stars \cite{2000ApJ...529L..25F}, the very
late thermal pulse (VLTP) in post-AGB stars (e.g., Sakurai's object,
\cite{2011ApJ...727...89H}) and the He-core flash in low-metallicity low-mass stars
(\cite{2010A&A...522L...6C}). 
The i-process in VLTP models with mixing assumptions motivated by 3D hydrodynamic simulations was found to
reproduce the observed abundance distribution in Sakurai's object \cite{2011ApJ...727...89H}. 
It may also explain the excess of $^{32}$S and anomalous Ti isotopic ratios in presolar A+B grains 
\cite{2013ApJ...776L..29F,2013ApJ...777L..27J}. 

\noindent In this paper, we investigate if the i-process nucleosynthesis can explain the observed 
abundance patterns in peculiar CEMP-r/s stars, such as CS 31062-050. We also briefly discuss the impact of nuclear physics 
uncertainties on the calculated i-process yields, as well as the results of 3D hydrodynamic
simulations of nuclear burning and mixing under conditions relevant to the i-process in Sakurai's object
recently reported in \cite{2014ApJ...792L...3H,2013arXiv1307.3821W}.



\section{A simple i-proces model for the CEMP-r/s stars} \label{sec_ipromodel}

\begin{figure}[!htbp]
\centering
\resizebox{8.8cm}{!}{\rotatebox{0}{\includegraphics{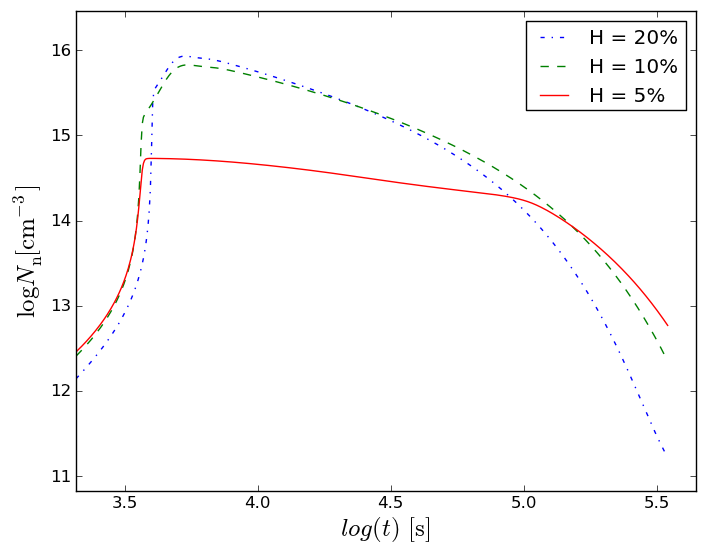}}}
\resizebox{9.5cm}{!}{\rotatebox{0}{\includegraphics{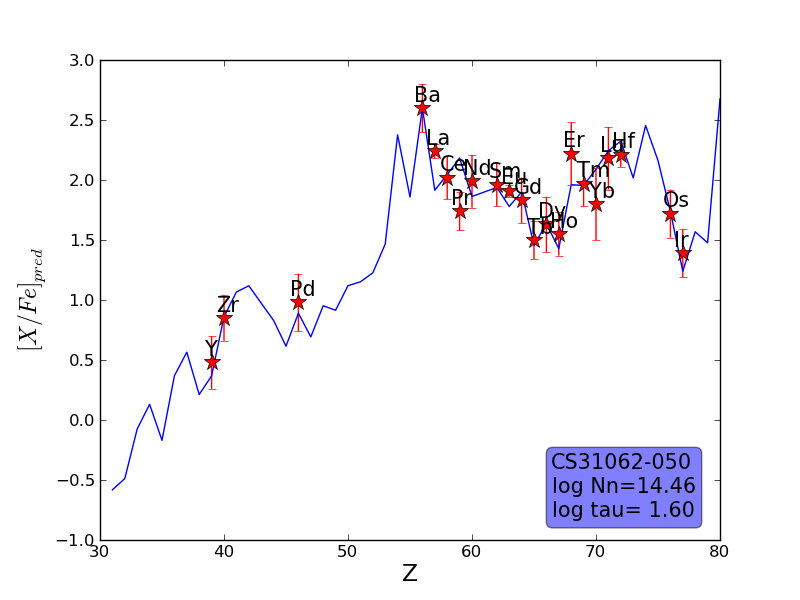}}}
\caption{
The evolution of the neutron density $N_\mathrm{n}$ with time for the initial H mass fractions
0.2, 0.1 and 0.05 (upper panel).
The best fit of the abundance pattern in the CEMP-r/s star CS 31062-050 with the i-process model 
(lower panel, with observational data from \cite{2004ApJ...605..462J}). 
Shown in the bottom-right corner are the fitted values of the neutron density $N_\mathrm{n}$ and
exposure (tau).}
\label{bisterzo12CS31062050_fit}
\end{figure}


\noindent We compare the abundance pattern in CS 31062-050 with results of our nucleosynthesis 
simulations of the i-process in one zone with constant temperature and density 
corresponding to the physical conditions in a He pulse-driven convective zone (PDCZ) from 1D stellar evolution models of AGB stars.
\noindent Our assumed values of $T = 2\times10^8\, \mathrm{K}$  and $\rho = 10^4\, \mathrm{g\,cm}^{-3}$ prevent
destruction of $^{13}$N via the (p,$\gamma$) channel, but they allow the neutron release via $^{13}$C($\alpha,n$). 
In the He PDCZ of a real star, with H entrained by convection from the surrounding H-rich envelope, 
the reactions $^{12}$C(p,$\gamma)^{13}$N and $^{13}$C($\alpha$,n)$^{16}$O are spatially separated,
the first occurring close to the upper and the second close to the lower convective boundary. 
$^{13}$N produced in the first reaction
decays to $^{13}$C while being transported down by convection.
The one-zone model  
reproduces the characteristic i-process neutron density ($N_\mathrm{n}\sim 10^{15} \mathrm{cm}^{-3}$) and neutron
exposures of $10\dots$ $50\ \mathrm{mbarn}^{-1}$. 
The initial abundances are taken from the 
solar abundance distribution \cite{1993oee..conf...15G}, with isotopic ratios from \cite{2003ApJ...591.1220L},
scaled down to the metallicity $Z=10^{-3}$. The
abundances of C and O are additionally modified, so that they are close to those found in the He PDCZ, i.e.
$X(^{12}\mathrm{C})=0.5$ and $X(^{16}\mathrm{O})=0.05$.
A specified fraction of H, assumed to be ingested from the H-rich envelope, is added to the mixture, with
the combined mass fraction of C+H being always set to 0.7.
The i-process nucleosynthesis is followed with the NuGrid single-zone PPN
code \cite{2008nuco.confE..23H}. 

\noindent During the first second, $^{12}$C is almost all transformed into $^{13}$N which then
decays into $^{13}$C on the timescale of $9.97\, \mathrm{min}$. 
After the decay of $^{13}$N, the reaction $^{13}$C($\alpha$,n)$^{16}$O releases neutrons with a high neutron density.
The upper panel of \fig\ref{bisterzo12CS31062050_fit} shows the evolution of the neutron density $N_\mathrm{n}$ with time for
different initial mass fractions of H.
As the simulation proceeds, the neutron exposure $\tau = \int_{0}^{t}N_\mathrm{n}v_T\,dt'$ ($v_T$ is the neutron thermal 
velocity) increases with time, and the abundance distribution is shifted to heavier elements. The model results shown in 
the lower panel of \fig\ref{bisterzo12CS31062050_fit} give the best representation of the abundance pattern in CS 31062-050. 
Its corresponding neutron density and exposure are indicated in the panel's bottom-right corner.

\begin{figure}[!htbp]
\centering
\resizebox{7.5cm}{!}{\rotatebox{0}{\includegraphics{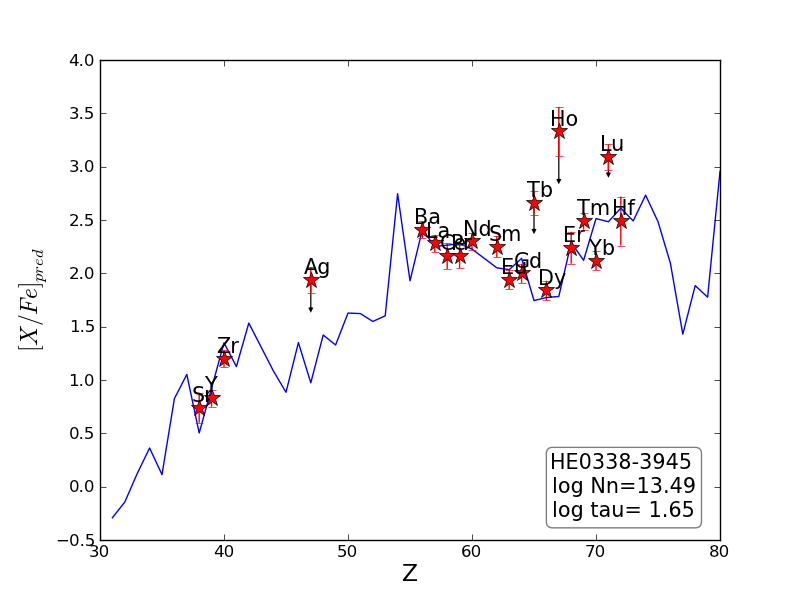}}}
\resizebox{7.5cm}{!}{\rotatebox{0}{\includegraphics{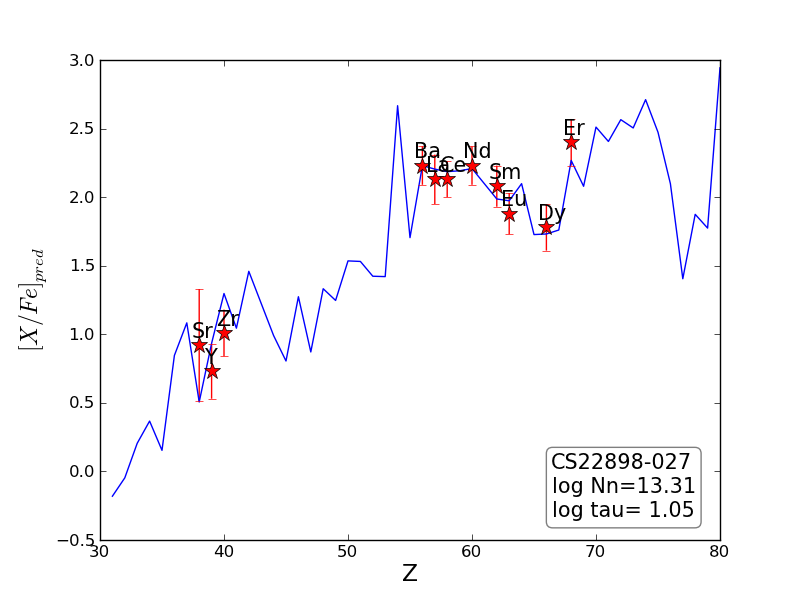}}}
\
\caption{Fit of the results of the one-zone i-process model to the abundance distributions in the CEMP-r/s stars 
HE 0338-3945 (left panel, with data from \cite{2006A&A...451..651J}) and 
CS 22898-027 (right panel, with data from \cite{2002ApJ...567.1166A}) for the indicated
neutron densities ($N_\mathrm{n}$) and exposures (tau).} 
\label{cempsr_examples}
\end{figure}

\noindent
Hydrodynamic simulations suggest that the i-process nucleosynthesis in the stellar environment is terminated 
when the energy released through the combustion of ingested H leads to a significant perturbation of 
the spherically symmetric convective shell structure. This self-quenching will depend on the exact configuration of 
the convective shells at the time of H-ingestion and can be expected to vary.
Therefore, a range of the termination
time is expected for the i-process, which motivates us to consider it as a free parameter in this simple model.
However, in spite of its simplicity, our one-zone i-process model reproduces surprisingly well
the entire heavy-element abundance distribution from Y to Ir within the observational errors in CS 31062-050 
(\fig\ref{bisterzo12CS31062050_fit}).
Specifically, this includes the large ratio of [hs/ls]\,$\approx 1$,  the [La/Eu] ratios between $0.0$ and $0.5$, which are 
between the ratios predicted for the s- and r-processes, as well as
high abundances in the Er-W region compared to those in the Os-Ir region, which are characteristic
signatures of the i-process nucleosynthesis at high neutron density, and which are all present 
in the abundance pattern of CS 31062-050. 
On the other hand, the one-zone model is not suited to correctly describe the evolution of the Fe group elements and Pb, 
because of its unrealistic depletion of the n-capture seed elements.
The observed abundance distributions in the CEMP-r/s stars HE 0338-3945 \cite{2006A&A...451..651J} and 
CS 22898-027 \cite{2002ApJ...567.1166A} 
are also well reproduced by the one-zone i-process model (\fig\ref{cempsr_examples}). 


\section{Nuclear physics uncertainties relevant to the i-process}
\begin{figure}[!htbp]
\centering
\resizebox{13.5cm}{!}{\rotatebox{0}{\includegraphics{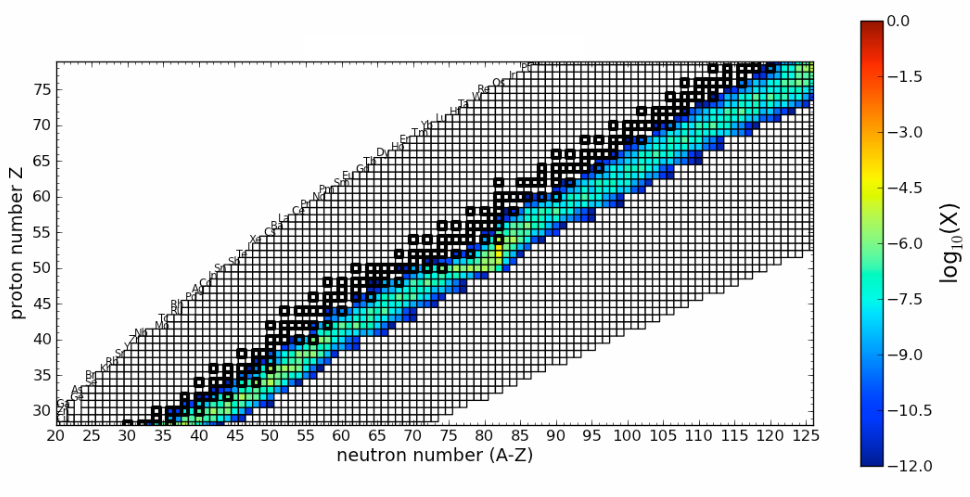}}}
\
\caption{The i-process nucleosynthesis path in the chart of isotopes with the color-coded mass fraction ($X$) of 
each species from our one-zone simulations.}
\label{bertolli13}
\end{figure}

\noindent When comparing predicted nucleosynthesis yields from the i-process with observations,
it is important to understand what and how strongly nuclear physics uncertainties can affect the calculated abundances of isotopes
along the i-process nucleosynthesis path, that goes four to five species
off the valley of stability, as shown in \fig\ref{bertolli13}.
The first i-process uncertainty analysis was done in \cite{2013arXiv1310.4578B},  where 
propagating systematic uncertainties of nuclear reaction cross sections from different theoretical models were investigated
for elements of the second s-process peak. The analysis considered Ba, La and Eu. 
The same i-process one-zone model described in Section \ref{sec_ipromodel}
was used. 
Fig. 6 in \cite{2013arXiv1310.4578B} shows changes of the [La/Eu] versus [Ba/La] ratio
for cases when nuclear reaction rates are estimated using
three different theoretical Hauser-Feshbach codes: NON-SMOKER \cite{2000nonsmoker.rauscher} (with rates from 
JINA REACLIB \cite{2010ApJS..189..240C}), 
CoH 3.3 \cite{2004proceedingKawano} and TALYS 1.4 \cite{2011talyskonig}.
The changes were found to be strongly model-dependent. 
The resulting [Ba/La] and [La/Eu] ratios varied, at least, by factors of $\sim 0.7$ dex and $\sim 0.3$ dex, respectively, 
between the models.
The conclusions are that the nuclear physics uncertainties strongly limit, at present, the predictive
power of i-process simulations \cite{2013arXiv1310.4578B} and that it would be very
important to further study the impact of nuclear physics uncertainties on the i-process element production at various sites.

\section{The i-process conditions in Sakurai's object}

\begin{figure}[!htbp]
\centering
\resizebox{7.0cm}{!}{\rotatebox{0}{\includegraphics{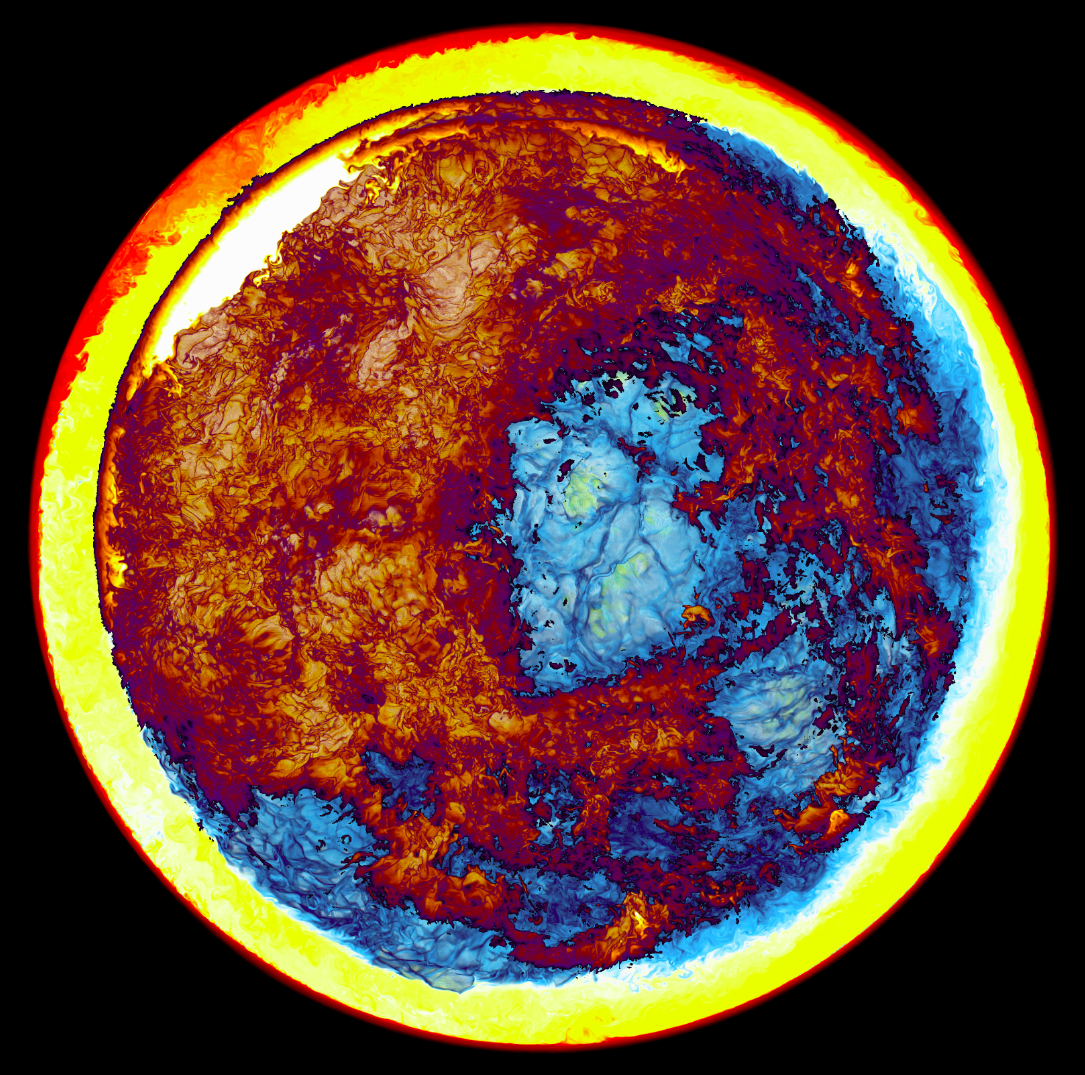}}}
\caption{A hemisphere with mixtures of entrained H-rich gas and He+C-rich gas of the He PDCZ.
The energy release rate from the burning of ingested H is shown in very dark blue, yellow, and white.}


\label{gosh}
\end{figure}

\noindent 
3D hydrodynamic simulations of mixing and H entrainment in the He PDCZ
of the post-AGB star Sakurai's object are now possible at fine enough
grids, so that the entrainment process is numerically converged
\cite{2013arXiv1307.3821W}.  Combustion of ingested protons in the
reaction $^{12}$C(p,$\gamma$)$^{13}$N and its energy feedback on
mixing was taken into account \cite{2014ApJ...792L...3H}.
Two runs with $768^3$ and $1536^3$ grid resolutions were performed and
showed quantitatively quite similar behaviour.  The continued H
ingestion led to its accumulation in the upper part of the convective
zone.  At some point, a large amount of H fuel ignited there and a
resulting burning pocket rose and collided with the stiff convective
boundary. The mass-conservation law then made the pocket to spread and
move along the boundary to the antipode, where the two fronts collided
causing a strong downdraft of the material, and the process repeated
several times.  That new mode of non-radial instabilities, coined by
the authors ``The Global Oscillation of Shell H-ingestion (GOSH)'',
first appeared after nearly 850 min in the $1536^3$ run. It is shown
in \fig\ref{gosh} at 1149 min
\cite{2014ApJ...792L...3H,2013arXiv1307.3821W}.
Such a GOSH event becomes more and more violent as it is repeated for about a dozen times in a row, leading      
to entrainment of entropy from the stable layer. 
In the run with $768^3$ grid resolution, a new upper convective boundary forms, while the 
run with $1536^3$ cells has not been followed to this point yet. 
It is not clear yet how the GOSH can be taken into account in 1D stellar evolution calculations.

\section{Conclusion}
\noindent Simple one-zone simulations of i-process nucleosynthesis
give yields that can fit almost all of the observed heavy-element
abundances from Y to Ir, within their observational errors, in some of the CEMP-r/s stars.  
The observed high enrichments with s- and r-process material in these stars
cannot be explained by the currently prevailing model that assumes a superposition of yields from different
s- and r-process sources.  These CEMP-r/s stars could therefore represent an i-process site.
Nuclear physics uncertainties are a main obstacle to obtaining reliable yields  
from the i-process nucleosynthesis calculations, as found by \cite{2013arXiv1310.4578B}.  
Mixing and H entrainment relevant to the i-process in the post-AGB star Sakurai's object have recently been investigated, which
has revealed a new mode of non-radial instabilities, the GOSH \cite{2014ApJ...792L...3H}.

\noindent {\bf Acknowledgments.} This material is based upon work supported by the National Science Foundation under 
Grant No. PHY-1430152 (JINA Center for the Evolution of the Elements).



\end{document}